%% file: main.tex
\documentclass[conference]{IEEEtran}

\usepackage[a4paper, total={184mm,239mm}]{geometry}

\input{mypackages.tex}

\input{mycommands.tex}

\input{mycolors.tex}

\colorlet{Iexcolor}{r}
\colorlet{VMcolor}{deepskyblue}
\colorlet{VGNAcolor}{darkMagenta}
\colorlet{VGKcolor}{b}

\graphicspath{{figures/}}    
    
\begin{document}

\title{
On the Excitability of 
Ultra-Low-Power
CMOS Analog Spiking Neurons
\thanks{This research was supported by the European Research Council (ERC) under the European Union’s Horizon AG research and innovation program (ERC-Synergy, SWIMS, 101119062) and the Research Projects “Thermodynamics of Circuits for Computation” and “Stochastic Modelling of Present and Future Nonlinear Dynamical Electronic Devices and Circuits” of the F.R.S.-FNRS.}
}


\author{
\IEEEauthorblockN{Léopold Van Brandt, Grégoire Brandsteert and Denis Flandre}
\IEEEauthorblockA{
        ICTEAM Institute, 
		UCLouvain, Louvain-la-Neuve, Belgium       
        \\
       {\tt leopold.vanbrandt@uclouvain.be}
      			 }
       }
       
\IEEEoverridecommandlockouts
\IEEEpubid{\makebox[\columnwidth]
{979-8-3315-7097-2/26/\$31.00~\copyright2026 IEEE \hfill}
\hspace{\columnsep}\makebox[\columnwidth]{ }}
       
\maketitle

\begin{abstract}
The excitability property of spiking neurons describes their capability to output an action potential as a real-time response to an input synaptic excitation current and is central to the event-based neuromorphic computing paradigm.
The spiking mechanism is analysed in a representative ultra-low-power analog neuron from the circuit literature.
Relying on conventional SPICE simulations compatible with industrial transistor compact models, we establish a excitation criterion, quantified either in terms of critical supplied charge or membrane potential threshold.
Only the latter is found intrinsic to the neuron, i.e. independent of the input stimulus.
Rigorous analysis of the nonlinear neuron dynamics provides insight but still needs to be explored further, as well as the effect of the intrinsic noise.
\end{abstract}


\section{Introduction}

Within the neuromorphic paradigm, spiking neurons are the fundamental building blocks of spiking neural networks.
Neuronal cells are massively replicated and interconnected through synapses to perform computation in real time~\cite{Chicca2014_PIEEE}.
A large variety of neuron architectures can be found~\cite{Indiveri2011}, either in CMOS technology (either analog~\cite{Chicca2014_PIEEE,Indiveri2011,Querlioz2013,Danneville2017} or digital~\cite{Frenkel2018}), or based on emerging devices such as volatile memristors~\cite{carboni2019stochastic,ascoli2025edge}.
There is an obvious tradeoff between the complexity of the architecture (i.e. required number of devices) that translates into higher power consumption, and the biophysical plausibility of the mathematical model of the neuron (which can be quantified by the number of behaviours of biological neurons that can be mimicked~\cite{izhikevich_which_2004}).
With this regard, the Morris-Lecar model~\cite{Morris1981}, that translates well to subthreshold MOS transistor physics, can be appreciated as a tradeoff between the basic integrate-and-fire~\cite{Indiveri2011} and the conductance-based Hodgkin-Huxley~\cite{Hodgkin1952}.

\begin{figure}[]
\centering
\captionsetup[subfigure]{singlelinecheck=off,justification=raggedright}
\captionsetup[subfigure]{skip=0pt}
\begin{subfigure}[t]{\linewidth}
\centering
\subcaption{
\vspace{-\baselineskip}
}
\label{fig_neuron}
\begin{tikzpicture}
\large
    \node[anchor=south west,inner sep=0] (image) at (0,0) {\includegraphics[width=0.8\linewidth]{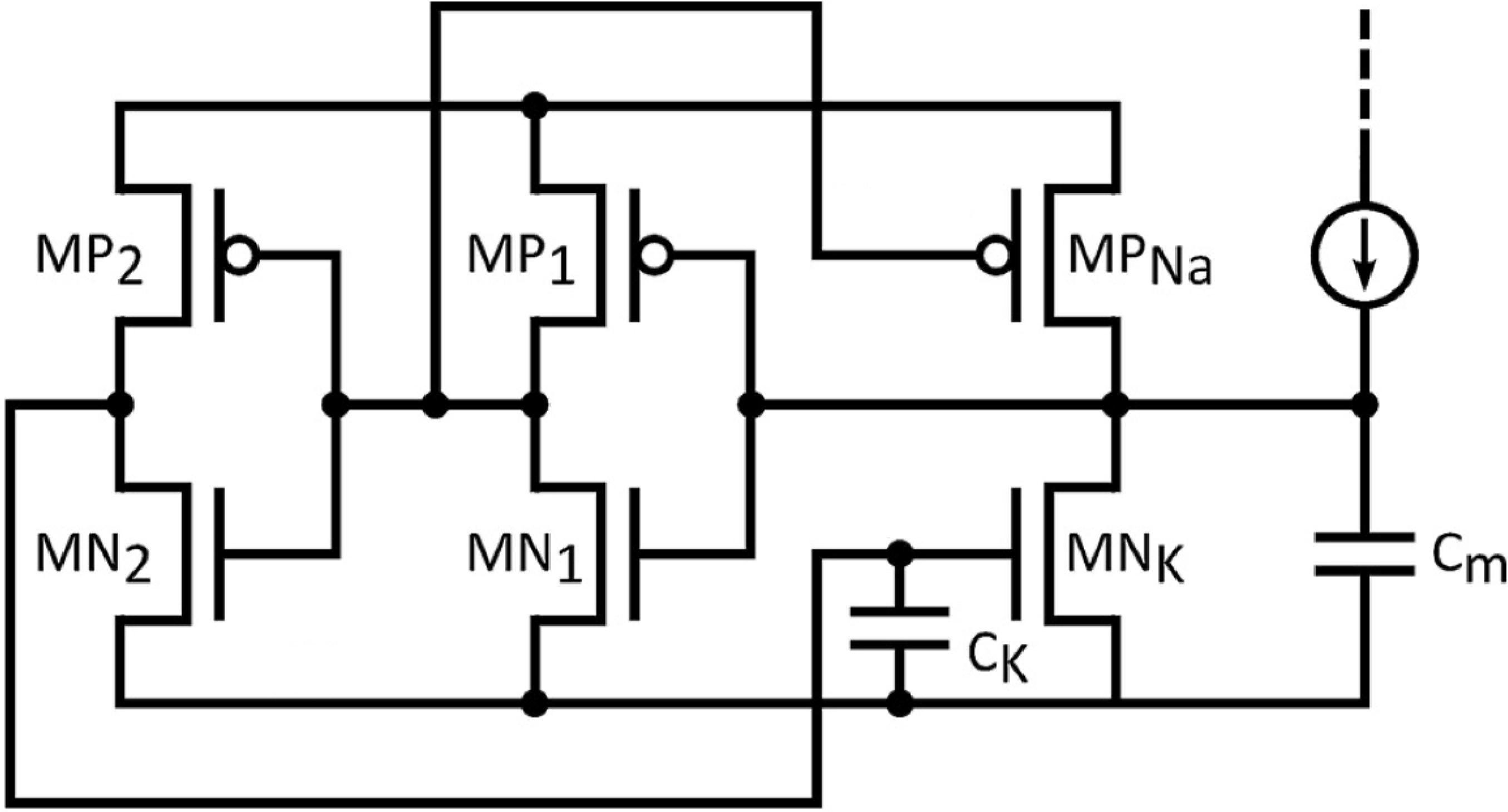}};
    \begin{scope}[x={(image.south east)},y={(image.north west)}]
    \node[anchor=west] at (0.925,0.75) {\small\color{Iexcolor}$\iex$};   
    \node[] at (0.905,0.5) {\large\color{VMcolor}$\bullet$}; 
    \node[anchor=west] at (0.915,0.50) {\small\color{VMcolor}$\vM$};
	\node[anchor=west] at (0.50,0.55) {\small\color{VMcolor}$\vM$};
	\node[] at (0.29,0.5) {\large\color{VGNAcolor}$\bullet$}; 
	\node[anchor=west] at (0.530,0.74) {\small\color{VGNAcolor}$\vGNA$};
	\node[anchor=west] at (0.220,0.425) {\small\color{VGNAcolor}$\vGNA$};
	\node[] at (0.080,0.5) {\large\color{VGKcolor}$\bullet$}; 
	\node[anchor=west] at (0.535,0.375) {\small\color{VGKcolor}$\vGK$};
	\node[anchor=west] at (-0.03,0.55) {\small\color{VGKcolor}$\vGK$};
	\node[anchor=west] at (0.16,0.925) {\small\color{k}$\VDD$};
	\node[anchor=west] at (0.16,0.185) {\small\color{k}$\SI{0}{\volt}$};
    \end{scope}
\end{tikzpicture}
\end{subfigure}

\vspace{+4mm}

\begin{subfigure}[t]{\linewidth}
\newcommand\myfontsize{\normalsize}
\newcommand\mytickfontsize{\footnotesize}
\myfontsize
\centering
\psfragscanon
\psfrag{t [us]}[cc][cc]{$t \, [\si{\micro\second}]$}
\psfrag{v(t) [mV]}[cc][cc]{$v(t) \, [\si{\milli\volt}]$}
\psfrag{iex [pA]}[cc][cc]{ $\textcolor{Iexcolor}{\iex \, [\si{\pico\ampere}]}$}
\psfrag{0}[cc][cc]{\mytickfontsize$0$}
\psfrag{5}[cc][cc]{\mytickfontsize$5$}
\psfrag{10}[cc][cc]{\mytickfontsize$10$}
\psfrag{15}[cc][cc]{\mytickfontsize$15$}
\psfrag{20}[cc][cc]{\mytickfontsize$20$}
\psfrag{50}[cc][cc]{\mytickfontsize$50$}
\psfrag{100}[cc][cc]{\mytickfontsize$100$}
\psfrag{150}[cc][cc]{\mytickfontsize$150$}
\psfrag{200}[cc][cc]{\mytickfontsize$200$}
\psfrag{0i}[cc][cc]{\mytickfontsize\color{Iexcolor}$0$}
\psfrag{40i}[cc][cc]{\mytickfontsize\color{Iexcolor}$40$}
\psfrag{vM}[tl][tl]{$\bm{\textcolor{VMcolor}{\vM}}$}
\psfrag{vGNA}[tl][tl]{$\textcolor{VGNAcolor}{\vGNA}$}
\psfrag{vGK}[tl][tl]{$\textcolor{VGKcolor}{\vGK}$}
\psfrag{O}[bc][bc]{0}
\psfrag{I}[bc][bc]{1}
\psfrag{II}[bc][bc]{2}
\psfrag{III}[bc][bc]{3}
\psfrag{vMres}[br][br]{\footnotesize$\textcolor{k}{\vMres}$}
\psfrag{vMth}[tl][tl]{\footnotesize$\textcolor{k}{\vMth}$}
\includegraphics[scale=1]{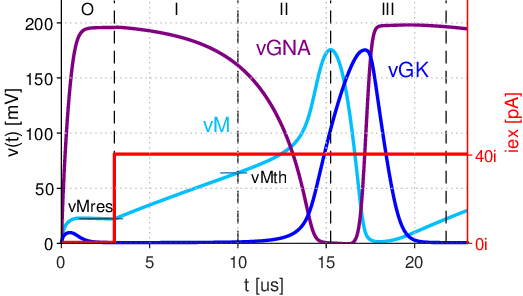}
\vspace{-5cm} 
\vspace{-2mm} 
\subcaption{}  
\label{fig_waveforms}
\end{subfigure}
\caption{
\subref{fig_neuron} Diagram of a CMOS analog spiking neuron~\cite{Sourikopoulos2017}.
\vspace{+1mm}
\newline
\subref{fig_waveforms} Waveforms at the different nodes of the circuit labelled in \subref{fig_neuron} under constant 
excitation current pulse
of amplitude $\Iex = \SI{40}{\pico\ampere}$.
Resting membrane potential is $\vMres \approx \SI{22}{\milli\volt}$ and threshold potential $\vMres \approx \SI{64}{\milli\volt}$.
\vspace{+1mm}
\newline
Case study: 
$\SI{65}{\nano\meter}$ CMOS technology design of~\cite{Sourikopoulos2017}, whose parameters are given in \cref{tab:param} 
and operating at ultra low \mbox{$\VDD = \SI{200}{\milli\volt}$} and $T = \SI{300}{\kelvin}$.
}
\label{fig_neuron_waveforms}


\vspace*{3mm}

\captionof{table}{
Transistor width $W$ and capacitances
of the neuron of \cref{fig_neuron}. All the transistor lengths are $L = \SI{65}{\nano\meter}$~\cite{Sourikopoulos2017}.
}
\label{tab:param}
{
\scriptsize
\begin{adjustbox}{center}
\begin{tabular}{cccccccc}
\toprule
MP\textsubscript{1} & MP\textsubscript{2} & MP\textsubscript{NA} & MN\textsubscript{1} & MN\textsubscript{2} & MN\textsubscript{K} & $\CK$ & $\CM$
\\
\midrule
$\SI{300}{\nano\meter}$ & $\SI{360}{\nano\meter}$ & $\SI{400}{\nano\meter}$ & $\SI{600}{\nano\meter}$ & $\SI{120}{\nano\meter}$ & $\SI{1.2}{\micro\meter}$ & $\SI{8}{\femto\farad}$ & $\SI{4}{\femto\farad}$
\\
\bottomrule
\end{tabular}
\end{adjustbox}
}
\vspace*{-2mm}
\end{figure}

An ultra-low-power CMOS design proposed by~\cite{Sourikopoulos2017} is shown in \cref{fig_neuron}.
Although relatively simple, the architecture has strong similarities (in terms of elementary blocks and operation) with adaptive exponential (leaky) integrate-and-fire neuron circuits~\cite{Chicca2014_PIEEE} as integrated in state-of-the-art neuromorphic processors~\cite{Greatorex2025}, and will therefore be used as case study.
While previous work~\cite{MIXDES2025} performed some preliminary analyses of the neuron reliability in presence of process variations, an insightful and quantitative spiking criterion remained to be established.
This connects to the notion of \emph{excitability} of 
neurons~\cite{izhikevich2000} that is central to the event-based analog neuromorphic 
computing paradigm~\cite{liu2014event_book}.
 Papers~\cite{Sepulchre2018,Sepulchre2025} investigated excitability based on theoretical 
neuron models only, in a control perspective.

The present work 
aims instead to
rely on conventional SPICE simulations that are not specific to a particular neuron architecture and moreover compatible with advanced industrial transistor compact models. 
The rest of this paper is structured as follows.
The operation of the neuron of \cref{fig_neuron}
is extensively described in \cref{section:Spiking Neuron Operation}. Excitability property and spiking criterion are addressed in \cref{section:Dynamics and Excitability of the Neuron}. \Cref{section:Conclusions and Perspectives} draws the conclusions and opens
perspectives for future work.



\section{Spiking Neuron Operation}
\label{section:Spiking Neuron Operation}


The operation of the neuron of \cref{fig_neuron}, leading to one action potential, that is one spike on the output $\vGK$, is illustrated in \cref{fig_waveforms} following successive phases:
\begin{itemize}
\item[0]
We assume cold initial conditions:
{\setlength{\belowdisplayskip}{0pt} \setlength{\belowdisplayshortskip}{0pt}
\setlength{\abovedisplayskip}{0pt} \setlength{\abovedisplayshortskip}{0pt}
 \[
  (\vM(0),\vGNA(0),\vGK(0)) = (\SI{0}{\volt},\SI{0}{\volt},\SI{0}{\volt})
  \text{,}
 \]}i.e. no initial residual charge in the circuit of \cref{fig_neuron}.
During \emph{power-up}, the supply voltage is turned from $0$ to $\VDD$ 
and the circuit state naturally evolves towards a stable steady state (equilibrium)~\cite{Vatajelu2016_SRAM_PUF}.
The stabilisation time, of a few $\si{\micro\second}$ for the design of \cref{tab:param}, is mostly determined by the inverter dynamics.
The membrane potential $\vM$ stabilises at its \emph{resting potential} $\vMres$.
\item[1]
Once the input synaptic excitation current $\iex$ is 
injected
and integrated over the membrane capacitance, $\vM$ increases quasi linearly, while the evolution of $\vGNA$ is dictated by inverter MN\textsubscript{1}-MP\textsubscript{1}.
\item[2]
After $\vM$ crosses a certain \emph{threshold} $\vMth$ (which will be quantitatively defined and extracted further), the decrease in $\vGNA$ becomes faster.
This
makes the pMOS transistor MP\textsubscript{NA} more and more conducting (yet in subthreshold regime for the studied ultra-low $\VDD$ design), which results in an exponential increase in $\vM$.
This, in turn, further reinforces the decrease in $\vGNA$: the neuron has entered in \emph{strong feedback mode}, 
similar to
SRAM state-transition dynamics~\cite{Zhang2006,TCAS2025}.
This positive feedback makes all the node voltages vary almost exponentially as can be observed in \cref{fig_waveforms}.
Notably, the quickly decreasing $\vGNA$ activates more and more the inverter MN\textsubscript{2}-MP\textsubscript{2}, which starts to generate 
an output spike on $\vGK$.
\item[3]
The transistor MP\textsubscript{K} becomes activated by the high $\vGK$ and inhibits the membrane potential, that is it lowers $\vM$ almost to the ground.
If, and only if, the excitation current is sustained, $\vM$ starts to rise again and the spiking cycle can resume from phase 1; otherwise the neuron returns to its stable equilibrium ($\vM$ at $\vMres$).
\end{itemize}

\section{Dynamics and Excitability of the Neuron}
\label{section:Dynamics and Excitability of the Neuron}


Strictly speaking, the neuron of \cref{fig_neuron} is a third-order system, 
whose state trajectory must be described by the triplet $(\vM(t),\vGNA(t),\vGK(t))$ and plotted in a three-dimensional (3D) state space. 
As pointed out in~\cite{Sepulchre2018}, the description of such nonlinear system dynamics in terms of trajectories is daunting. 
Therefore, complex neurons are more conveniently characterised in terms of \emph{excitability}, that is an intrinsic input-output property that does not necessarily require the knowledge of the full internal state model~\cite{Sepulchre2018,Sepulchre2022_PIEEE,Sepulchre2025}.

\subsection{Excitation Threshold and Critical Charge}

\begin{figure}[]
\captionsetup[subfigure]{singlelinecheck=off,justification=raggedright}
\newcommand\myfontsize{\normalsize}
\newcommand\mytickfontsize{\small}
\captionsetup[subfigure]{skip=0pt}
\begin{subfigure}[t]{\linewidth}
\renewcommand\myfontsize{\small}
\renewcommand\mytickfontsize{\footnotesize}
\myfontsize
\centering
\subcaption{
\vspace{-\baselineskip}
}
\label{fig_excitability}
\psfragscanon
\psfrag{Iex [pA]}[cc][cc]{$\Iex \, [\si{\pico\ampere}]$}
\psfrag{pw [us]}[cc][cc]{$\pw \, [\si{\micro\second}]$}
\psfrag{0}[cc][cc]{\mytickfontsize$0$}
\psfrag{10}[cc][cc]{\mytickfontsize$10$}
\psfrag{20}[cc][cc]{\mytickfontsize$20$}
\psfrag{30}[cc][cc]{\mytickfontsize$30$}
\psfrag{40}[cc][cc]{\mytickfontsize$40$}
\psfrag{50}[cc][cc]{\mytickfontsize$50$}
\psfrag{60}[cc][cc]{\mytickfontsize$60$}
\psfrag{70}[cc][cc]{\mytickfontsize$70$}
\psfrag{80}[cc][cc]{\mytickfontsize$80$}
\psfrag{90}[cc][cc]{\mytickfontsize$90$}
\psfrag{100}[cc][cc]{\mytickfontsize$100$}
\psfrag{(b)}[bc][bc]{(b)}
\psfrag{Figure 1(b)}[bc][bc]{\cref{fig_waveforms}}
\psfrag{(c)}[tc][tc]{(c)}
\psfrag{pwcrit(Iex)}[bc][bc]{\color{mybrown}$\pwcrit(\Iex)$}
\includegraphics[scale=1]{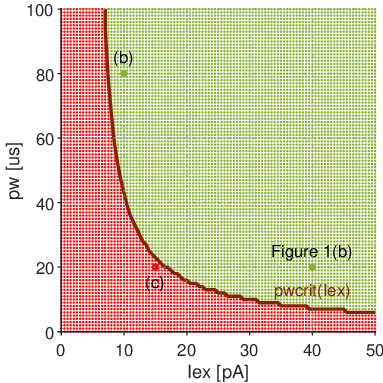}
\end{subfigure}

\vspace{+2mm}

\begin{subfigure}[t]{\linewidth}
\renewcommand\myfontsize{\small}
\renewcommand\mytickfontsize{\scriptsize}
\myfontsize
\centering
\subcaption{$\Iex = \SI{10}{\pico\ampere}$, $\pw = \SI{80}{\micro\second}$ (one spike)}
\label{fig_waveforms_Iex_10pA_pw_80us}
\psfragscanon
\psfrag{t [us]}[cc][cc]{$t \, [\si{\micro\second}]$}
\psfrag{v(t) [mV]}[cc][cc]{$v(t) \, [\si{\milli\volt}]$}
\psfrag{iex [pA]}[cc][cc]{ $\textcolor{Iexcolor}{\iex \, [\si{\pico\ampere}]}$}
\psfrag{0}[cc][cc]{\mytickfontsize$0$}
\psfrag{20}[cc][cc]{\mytickfontsize$20$}
\psfrag{40}[cc][cc]{\mytickfontsize$40$}
\psfrag{60}[cc][cc]{\mytickfontsize$60$}
\psfrag{80}[cc][cc]{\mytickfontsize$80$}
\psfrag{100}[cc][cc]{\mytickfontsize$100$}
\psfrag{200}[cc][cc]{\mytickfontsize$200$}
\psfrag{0i}[cc][cc]{\mytickfontsize\color{Iexcolor}$0$}
\psfrag{10i}[cc][cc]{\mytickfontsize\color{Iexcolor}$10$}
\psfrag{vM}[tl][tl]{$\bm{\textcolor{VMcolor}{\vM}}$}
\psfrag{vGK}[tl][tl]{$\textcolor{VGKcolor}{\vGK}$}
\psfrag{vMth}[tc][tc]{\footnotesize$\textcolor{gray}{\vMth}$}
\includegraphics[scale=1]{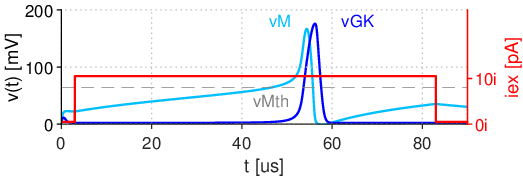}
\end{subfigure}

\vspace{+1mm}

\begin{subfigure}[t]{\linewidth}
\renewcommand\myfontsize{\small}
\renewcommand\mytickfontsize{\scriptsize}
\myfontsize
\centering
\subcaption{$\Iex = \SI{15}{\pico\ampere}$, $\pw = \SI{20}{\micro\second}$ (no spike)}
\label{fig_waveforms_Iex_15pA_pw_20us}
\psfragscanon
\psfrag{t [us]}[cc][cc]{$t \, [\si{\micro\second}]$}
\psfrag{v(t) [mV]}[cc][cc]{$v(t) \, [\si{\milli\volt}]$}
\psfrag{iex [pA]}[cc][cc]{ $\textcolor{Iexcolor}{\iex \, [\si{\pico\ampere}]}$}
\psfrag{0}[cc][cc]{\mytickfontsize$0$}
\psfrag{20}[cc][cc]{\mytickfontsize$20$}
\psfrag{40}[cc][cc]{\mytickfontsize$40$}
\psfrag{60}[cc][cc]{\mytickfontsize$60$}
\psfrag{80}[cc][cc]{\mytickfontsize$80$}
\psfrag{100}[cc][cc]{\mytickfontsize$100$}
\psfrag{200}[cc][cc]{\mytickfontsize$200$}
\psfrag{0i}[cc][cc]{\mytickfontsize\color{Iexcolor}$0$}
\psfrag{15i}[cc][cc]{\mytickfontsize\color{Iexcolor}$15$}
\psfrag{vM}[tl][tl]{$\bm{\textcolor{VMcolor}{\vM}}$}
\psfrag{vGK}[bl][bl]{$\textcolor{VGKcolor}{\vGK}$}
\psfrag{vMth}[tc][tc]{\footnotesize$\textcolor{gray}{\vMth}$}
\includegraphics[scale=1]{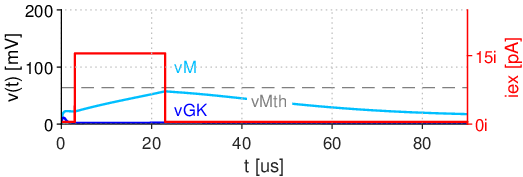}
\end{subfigure}
\caption{
\subref{fig_excitability}
Threshold property of the excitable neuron of \cref{fig_neuron} (plot suggested by~\cite{Sepulchre2018,Sepulchre2025}), extracted from deterministic SPICE transient simulations with a sweep on excitation current pulse parameters $(\Iex,\pw)$,
resp.
amplitude and 
duration. 
The green region corresponds to input pulses $\iex$ 
that lead to an action potential, i.e. a spike on output $\vGK$. 
Low-energy pulses 
located in
the red region fail to active the 
spiking mechanism of the circuit.
\vspace{+1mm}
\newline
\subref{fig_waveforms_Iex_10pA_pw_80us} \subref{fig_waveforms_Iex_15pA_pw_20us}
Membrane and action potential waveforms of two illustrative cases marked by dots in \subref{fig_excitability}, in addition to the case shown in \cref{fig_waveforms}. For the pulse \subref{fig_waveforms_Iex_10pA_pw_80us}, the membrane potential crosses the threshold $\vMth$ and the neuron fires a spike.
The pulse \subref{fig_waveforms_Iex_15pA_pw_20us} does not supply a sufficient amount of energy.
\vspace{-3mm}
}
\label{fig_excitability_waveforms}
\end{figure}

In \cref{fig_excitability}, we have assessed the excitability of the neuron of \cref{fig_neuron} through SPICE
 \textcolor{h}{noiseless}
transient simulations.
Starting from cold initial conditions as described in \cref{section:Spiking Neuron Operation}, we have injected synaptic current pulses (like the one 
shown in \cref{fig_waveforms}) sweeping amplitudes $\Iex$ and durations $\pw$.
For each case, we observe the 
output 
potential $\vGK$: the neuron has been successfully excited if one output spike is recorded (green region in \cref{fig_excitability}). Examples are \cref{fig_waveforms} and 
\cref{fig_waveforms_Iex_10pA_pw_80us}: a small-amplitude pulse 
triggers
a spike if it is sustained for a sufficient duration.
Similar criterion was established for 
a state transition to occur 
in SRAM bitcells subject to deterministic pulse disturbances~\cite{Zhang2006} or random telegraph noise~\cite{SSE2023}.
\Cref{fig_waveforms_Iex_15pA_pw_20us} shows one instance
of pulse that is too low and short to trigger the spiking mechanism, thereby located in the red region of  \cref{fig_excitability}.
In all, observation suggests to 
determine
a 
threshold
criterion
that would take into account both the amplitude \emph{and} the duration of the injected pulse.

For a given amplitude $\Iex$ of the excitation current pulse, the threshold boundary (in brown line in \cref{fig_excitability}) provides the minimum duration of the pulse, denoted $\pwcrit(\Iex)$, needed to generate a spike
on $\vGK$.
Conversely, a given $\pw$ imposes a minimum $\Iex$.
It can be 
noticed
that the integral of $\iex(t)$, i.e. the product of $\Iex$ and $\pw$ yields a \emph{critical charge}, i.e. the minimum amount of charge that must be supplied to excite the neuron~\cite{Sepulchre2018}:
{
\setlength{\abovedisplayskip}{2pt} \setlength{\abovedisplayshortskip}{2pt}
\begin{equation}
\label{eq:DQth}
\DQth(\Iex) = \int_{t_0}^{t_0+\pwcrit} \iex(t) \,\diff{t} = \Iex \cdot \pwcrit
\text{.}
\end{equation}}This allows to extend the analysis to other types of synaptic current than a rectangular pulse, like a spike train as will be exemplified in \cref{subsection:Always-Spiking Configuration}.

However, our numerical investigation reveals that the critical charge varies with $\Iex$ and hence is not a unique quantity, which the notation $\DQth(\Iex)$ in \eqref{eq:DQth} reflects. 
The excitation boundary in \cref{fig_excitability} is indeed not an isoenergetic curve~\cite{Sepulchre2025}.
Since the critical charge can depend on the input stimulus, it is not an intrinsic property of the neuron.
This non-trivial dependency comes from the nonlinear dynamics of the neuron and the complexity of its responses to different stimuli.
The difficulty of proposing a simple and general definition of the threshold of excitable systems was recently 
pointed out
by~\cite{Sepulchre2025}.

\subsection{Membrane Potential Threshold as Spiking Criterion}

The previous section has highlighted the difficulty of identifying a unique excitability threshold in terms of charge or energy only based on the input-output response of the system (black-box approach).
This issue gets solved if internal states of the system are observable and their dynamical evolution can be modelled, as is the case here with the detailed knowledge of the circuit architecture (\cref{fig_neuron}) and the possibility to probe any signal in SPICE.

\begin{figure}[]
\newcommand\myfontsize{\small}
\newcommand\mytickfontsize{\footnotesize}
\myfontsize
\centering
\psfragscanon
\psfrag{t [us]}[cc][cc]{$t \, [\si{\micro\second}]$}
\psfrag{vM(t) [mV]}[cc][cc]{$\textcolor{VMcolor}{\vM(t) \, [\si{\milli\volt}]}$}
\psfrag{CMtot [F]}[cc][cc]{$\textcolor{m}{\CMtot \, [\si{\femto\farad}]}$}
\psfrag{0}[cc][cc]{\mytickfontsize$0$}
\psfrag{2}[cc][cc]{\mytickfontsize$2$}
\psfrag{4}[cc][cc]{\mytickfontsize$4$}
\psfrag{6}[cc][cc]{\mytickfontsize$6$}
\psfrag{8}[cc][cc]{\mytickfontsize$8$}
\psfrag{10}[cc][cc]{\mytickfontsize$10$}
\psfrag{12}[cc][cc]{\mytickfontsize$12$}
\psfrag{0v}[cr][cr]{\mytickfontsize\color{VMcolor}\mytickfontsize$0$}
\psfrag{20v}[cr][cr]{\mytickfontsize\color{VMcolor}\mytickfontsize$20$}
\psfrag{40v}[cr][cr]{\mytickfontsize\color{VMcolor}\mytickfontsize$40$}
\psfrag{60v}[cr][cr]{\mytickfontsize\color{VMcolor}\mytickfontsize$60$}
\psfrag{80v}[cr][cr]{\mytickfontsize\color{VMcolor}\mytickfontsize$80$}
\psfrag{Cm}[cl][cl]{\mytickfontsize\color{m}$\CM$}
\psfrag{6f}[cl][cl]{\mytickfontsize\color{m}$\SI{6}{\femto\relax}$}
\psfrag{8f}[cl][cl]{\mytickfontsize\color{m}$\SI{8}{\femto\relax}$}
\psfrag{vM}[tl][tl]{$\bm{\textcolor{VMcolor}{\vM}}$}
\psfrag{vGNA}[tl][tl]{$\textcolor{VGNAcolor}{\vGNA}$}
\psfrag{vGK}[tl][tl]{$\textcolor{VGKcolor}{\vGK}$}
\psfrag{O}[tc][tc]{0}
\psfrag{I}[tc][tc]{1}
\psfrag{II}[tc][tc]{2}
\psfrag{vMres}[br][br]{\footnotesize$\textcolor{k}{\vMres}$}
\psfrag{vMth}[br][br]{\footnotesize$\textcolor{k}{\vMth}$}
\psfrag{DvMth}[cr][cr]{\footnotesize$\textcolor{k}{\DvMth}$}
\psfrag{CMtotav}[tc][tc]{\small\textcolor{violet}{$\CMtotav$}}
\psfrag{iex}[tr][tr]{\textcolor{Iexcolor}{$\iex$}}
\psfrag{pwcrit}[tc][tc]{\small\textcolor{mybrown}{$\pwcrit(\Iex)$}}
\includegraphics[scale=1]{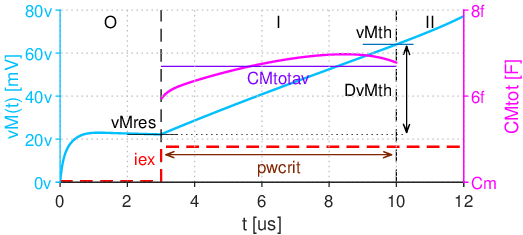}
\vspace{-2mm}
\caption{Definition of the threshold potential $\vMth$ as the value reached by $\vM$ when excited by a current pulse of duration $\pwcrit(\Iex)$ provided by \cref{fig_excitability} for some given amplitude $\Iex$.
While $\Iex = \SI{40}{\pico\ampere}$ was arbitrarily chosen for illustration (as in \cref{fig_waveforms}), the threshold $\vMth$ is an intrinsic property of the neuron and does not depend on $\Iex$.
The extraction of the nonlinear effective total membrane capacitance $\CMtot$ according to the definition \eqref{eq:CMtot} is also shown; 
$\CMtotav$ denotes the 
averaged effective capacitance over the phase 1 during which $\vM$ rises from $\vMres$ to $\vMth$.
\vspace{-3mm}
}
\label{fig_Cm}
\end{figure}

For all the bordelines cases of \cref{fig_excitability}, we have observed that the \emph{spiking criterion} is that the membrane potential exceeds a certain threshold $\vMth$ that is indicated in \cref{fig_waveforms,fig_waveforms_Iex_10pA_pw_80us,fig_waveforms_Iex_15pA_pw_20us}.
This is nothing but the expected natural behaviour of any 
integrate-and-fire neuron~\cite{Indiveri2011,Chicca2014_PIEEE}.
Thus, an output spike is 
fired
when the synaptic excitation current supplies a sufficient amount of energy to make $\vM$ rise from $\vMres$ to $\vMth$ (the phase 1 described in \cref{section:Spiking Neuron Operation}).
The potential difference will be denoted $\DvMth =  \vMth - \vMres$. All the important quantities are summarised in \cref{fig_Cm}.
 It would be insightful to relate it 
to the critical charge \eqref{eq:DQth} 
in order to unify the criteria. 

\subsubsection{Effective Total Membrane Capacitance}

The application of Kirchhoff's current law at node $\vM$ (see \cref{fig_neuron}) yields
\begin{equation}
\label{eq:KCL}
\iex(t) = \CM  \dv{\vM(t)}{t}
- i_{\mathrm{D,Na}}(t) + i_{\mathrm{D,K}}(t)
+ i_{\mathrm{in,Inv1}}(t)
\text{.}
\end{equation}
$i_{\mathrm{in,Inv1}}(t)$ is a pure dynamic current in the input gates capacitances of inverter MN\textsubscript{1}-MP\textsubscript{1}.
However, the drain currents of the transistors MP\textsubscript{Na} and MN\textsubscript{K}, $i_{\mathrm{D,Na}}(t)$ and $i_{\mathrm{D,K}}(t)$, have both quasi-static and dynamic components.
The quasi-static components 
are the leakage currents of the transistors, which are small in phase 1 when $\vGNA$ remains close to $\VDD$ and $\VGK$ is still almost at $\SI{0}{\volt}$ (see \cref{fig_waveforms}) but never rigorously zero.

We aim to extract an \emph{effective} total membrane capacitance (possibly nonlinear) such that
{
\setlength{\abovedisplayskip}{0pt} \setlength{\abovedisplayshortskip}{0pt}
\begin{equation}
\label{eq:CMtot}
\iex(t)
\equiv \CMtot(t) \, \dv{\vM(t)}{t}
\text{.}
\end{equation}}The $\CMtot$ defined by \eqref{eq:CMtot} is said effective because 
it encompasses the implemented $\CM$ (seen in \cref{fig_neuron}), the contribution of all intrinsic MOS and parasitic interconnect capacitances connected to the node $\vM$, as well as the effect of leakage currents discussed above.
Subsequently, we
extract the 
nonlinear 
$\CMtot(t)$ from SPICE transient simulation of the circuit (\cref{fig_Cm}) excited by a constant current pulse $\iex(t) = \Iex$.
The $\CMtot(t) = \Iex / (\diff{\vM}(t)/\diff{t})$, obtained from \eqref{eq:CMtot} in phase 1,
is plotted in magenta in \cref{fig_Cm}.
The non-constancy of $\CMtot(t)$ reflects the weak nonlinearity of $\vM(t)$.
It should be noticed that reducing the $\CMtot$ to $\CM$ 
(as given in \cref{tab:param}) 
only would be a coarse approximation.



Finally, the combination of \eqref{eq:DQth} and \eqref{eq:CMtot} 
consistently
relates the critical supplied charge to the membrane potential threshold:
\begin{equation}
\begin{aligned}[b]
\label{eq:DQth = CMtotav.DvMth}
\DQth = \CMtotav \cdot \DvMth
\text{,}
\end{aligned}
\end{equation}
where the averaged effective total membrane capacitance (thin violet line in \cref{fig_Cm}) has been introduced:
\begin{equation}
\label{eq:CMtotav}
\CMtotav (\Iex) \equiv \frac{1}{\DvMth} \int_{\vMres}^{\vMth} \CMtot(\vM) \, \diff{\vM}
\text{.}
\end{equation}

\subsubsection{Neuron Dynamics in Two-Dimensional State Subspace}

\begin{figure}[]
\newcommand\myfontsize{\normalsize}
\newcommand\mytickfontsize{\footnotesize}
\myfontsize
\centering
\psfragscanon
\psfrag{vM [mV]}[cc][cc]{$\vM \, [\si{\milli\volt}]$}
\psfrag{vGNA [mV]}[cc][cc]{$\vGNA \, [\si{\milli\volt}]$}
\psfrag{0}[cc][cc]{\mytickfontsize$0$}
\psfrag{50}[cc][cc]{\mytickfontsize$50$}
\psfrag{100}[cc][cc]{\mytickfontsize$100$}
\psfrag{150}[cc][cc]{\mytickfontsize$150$}
\psfrag{200}[cc][cc]{\mytickfontsize$200$}
\psfrag{O}[cl][cl]{\small\color{gray}0}
\psfrag{I}[cl][cl]{\small\color{VMcolor}1}
\psfrag{II}[cl][cl]{\small\color{VGKcolor}2}
\psfrag{III}[cl][cl]{\small\color{VGKcolor}3}
\psfrag{vMres}[br][br]{\footnotesize$\textcolor{k}{\vMres}$}
\psfrag{vMth}[bl][bl]{\footnotesize$\textcolor{k}{\vMth}$}
\psfrag{DvMth}[bc][bc]{\footnotesize$\textcolor{k}{\DvMth}$}
\psfrag{VGNA = f1(VM)}[cl][cl]{\color{mybrown}\footnotesize$\VGNA = \ffi(\VM)$}
\psfrag{VM = fNA(VGNA)}[cr][cr]{\color{mybrown}\footnotesize$\VM = \fNA(\VGNA)$}
\includegraphics[scale=1]{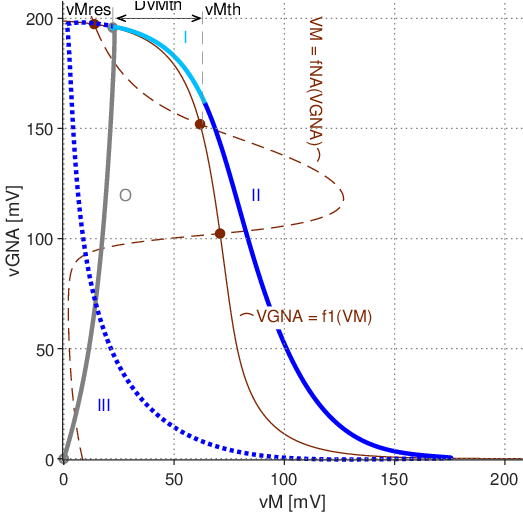}
\vspace{-2mm}
\caption{Trajectory in the 
2D
state subspace $(\vM,\vGNA)$, corresponding to the simulated waveforms and operation phases of \cref{fig_waveforms}. 
Membrane potential $\vMres$ and the threshold potential $\vMres$ are indicated.
The quasi-static voltage transfer characteristics 
$\VGNA = \ffi(\VM)$
(inverter MN\textsubscript{1}-MP\textsubscript{1})
and $\VM = \fNA(\VGNA)$
(transistor MP\textsubscript{Na} with a feedback effect of the whole circuit) are shown in brown lines, resp. full and dashed.
\vspace{-3mm}
}
\label{fig_2D_space}
\end{figure}

We conclude this section by explaining the found empirical value of the membrane potential threshold $\vMth$ with an analysis of the neuron dynamics.
\Cref{fig_2D_space} shows the two-dimensional (2D) state subspace of the neuron, where only the state variables $(\vM,\vGNA)$ are considered.
This 2D subspace can be regarded as a projection of the full 3D state space, meaning that the dynamics are approximated.
The state trajectory $(\vM(t),\vGNA(t))$ corresponding to the simulation of \cref{fig_waveforms} is depicted, along with the different phases annotated.
One can see the \emph{limit cycle}, i.e. the closed and periodic orbit 
followed when the neuron operates in always-spiking or oscillating configuration (when a sufficient large constant $\Iex$ or a continuous spike train is applied as synaptic current in \cref{fig_neuron} as will be shown in \cref{subsection:Always-Spiking Configuration}).

The quasi-static voltage transfer characteristics (VTCs)
$\VGNA = \ffi(\VM)$ and $\VM = \fNA(\VGNA)$, obtained by sweeping a DC voltage source introduced on purpose and applied at a node (resp. $\vM$ and $\vGNA$) of \cref{fig_neuron} are also shown (in brown) in \cref{fig_2D_space}. 
Akin to the butterfly plot for CMOS SRAM bitcells~\cite{TCAS2025}, the intersections of these curves determine the steady states (stable \emph{or} unstable) of the circuit.
The top-left brown point is found to 
underestimate $\vMres$ (from which the trajectory evolves in the beginning of phase 1, in sky blue): 
the error could result from
the current sunk by the DC voltage source (which alters the original circuit behaviour) and the projection 
from 3D to 2D. 
The middle intersection of the VTCs is an \emph{unstable} steady state whose $x$-location closely confirms the $\vMth$ found earlier from
SPICE 
transient simulations.
While deeper and more quantitative analyses of the neuron dynamics\textcolor{h}{~\cite{Izhikevich2007_book}} are left for future work, we can 
hypothesise
that this threshold point lies on a \emph{separatrix}~\cite{TCAS2025}.
Triggering a spike requires to excite the neuron so that its state $(\vM(t),\vGNA(t))$ crosses the separatrix to escape from the attraction of the resting state, which quantitatively explains the role of 
$\vMth$. 



%
%

\subsection{Always-Spiking Configuration}
\label{subsection:Always-Spiking Configuration}

\begin{figure}[]
\captionsetup[subfigure]{singlelinecheck=off,justification=raggedright}
\newcommand\myfontsize{\small}
\newcommand\mytickfontsize{\footnotesize}
\captionsetup[subfigure]{skip=0pt}
\begin{subfigure}[t]{\linewidth}
\myfontsize
\centering
\subcaption{Continuous pulse $\Iex = \SI{40}{\pico\ampere}$}
\vspace{+1mm}

\label{fig_waveforms_long_pulse}
\psfragscanon
\psfrag{t [us]}[cc][cc]{$t \, [\si{\micro\second}]$}
\psfrag{v(t) [mV]}[cc][cc]{$v(t) \, [\si{\milli\volt}]$}
\psfrag{iex [pA]}[cc][cc]{ $\textcolor{Iexcolor}{\iex \, [\si{\pico\ampere}]}$}
\psfrag{0}[cc][cc]{\mytickfontsize$0$}
\psfrag{50}[cc][cc]{\mytickfontsize$50$}
\psfrag{100}[cc][cc]{\mytickfontsize$100$}
\psfrag{150}[cc][cc]{\mytickfontsize$150$}
\psfrag{200}[cc][cc]{\mytickfontsize$200$}
\psfrag{0i}[cl][cl]{\mytickfontsize\color{Iexcolor}$0$}
\psfrag{40i}[cl][cl]{\mytickfontsize\color{Iexcolor}$40$}
\psfrag{vM}[tl][tl]{$\bm{\textcolor{VMcolor}{\vM}}$}
\psfrag{vGK}[tl][tl]{$\textcolor{VGKcolor}{\vGK}$}
\psfrag{vMth}[tc][tc]{\footnotesize$\textcolor{gray}{\vMth}$}
\includegraphics[scale=1]{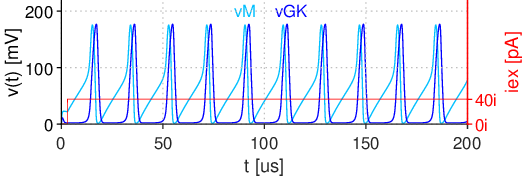}
\end{subfigure}

\vspace{+2mm}

\begin{subfigure}[t]{\linewidth}
\myfontsize
\centering
\subcaption{Spike train excitation with $\Iex = \SI{100}{\pico\ampere}$, $\pw = \SI{2}{\micro\second}$, 
$\DTq = \SI{5}{\micro\second}$
}
\label{fig_waveforms_spike_train}
\vspace{+1mm}
\psfragscanon
\psfrag{t [us]}[cc][cc]{$t \, [\si{\micro\second}]$}
\psfrag{v(t) [mV]}[cc][cc]{$v(t) \, [\si{\milli\volt}]$}
\psfrag{iex [pA]}[cc][cc]{ $\textcolor{Iexcolor}{\iex \, [\si{\pico\ampere}]}$}
\psfrag{0}[cc][cc]{\mytickfontsize$0$}
\psfrag{50}[cc][cc]{\mytickfontsize$50$}
\psfrag{100}[cc][cc]{\mytickfontsize$100$}
\psfrag{150}[cc][cc]{\mytickfontsize$150$}
\psfrag{200}[cc][cc]{\mytickfontsize$200$}
\psfrag{0i}[cl][cl]{\mytickfontsize\color{Iexcolor}$0$}
\psfrag{40i}[cl][cl]{\mytickfontsize\color{Iexcolor}$100$}
\psfrag{vM}[tl][tl]{$\bm{\textcolor{VMcolor}{\vM}}$}
\psfrag{vGK}[bl][bl]{$\textcolor{VGKcolor}{\vGK}$}
\psfrag{vMth}[tc][tc]{\footnotesize$\textcolor{gray}{\vMth}$}
\includegraphics[scale=1]{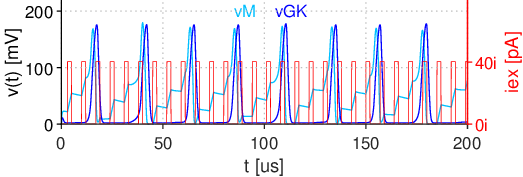}
\end{subfigure}
\vspace{-1mm}
\caption{
Action potential train (output spike train) sustained by \subref{fig_waveforms_long_pulse} a constant synaptic excitation current pulse; 
\subref{fig_waveforms_spike_train} a synaptic current spike train. 
\vspace{+1mm}
\newline
Illustrated case: same neuron design as \cref{fig_neuron_waveforms} and \cref{tab:param}.
\vspace{-3mm}
}
\label{fig_waveforms_long_pulse_spike_train}
\end{figure}

So far, we have focused on the mechanism for generating a single action potential under a constant synaptic excitation current. The critical supplied charge or the membrane potential threshold were established as spiking criterion.

As can be seen in \cref{fig_waveforms_long_pulse}, sustaining the synaptic current pulse yields an \emph{action potential train}, i.e. an output spike train on $\vGK$.
The output spike frequency obviously increases with $\Iex$, however nonlinearly as shown in previous works~\cite{Sourikopoulos2017,MIXDES2025}.

\Cref{fig_waveforms_spike_train} depicts the waveforms when the neuron is excited by an \emph{input synaptic current spike train}, which emulates a more biologically plausible stimulus~\cite{Koch2004} moreover relevant for event-based neuromorphic computing systems~\cite{liu2014event_book}.
Each individual spike is \emph{integrated} on the membrane capacitance, thereby supplying a certain amount of charge and gradually increases $\vM$.
During the \emph{refractory period}~\cite{Indiveri2011}, that is the quiescient period $\DTq$ between two input spikes, $\vM$ can be observed to decrease: this is the manifestation of the \emph{leaky} property of the neuron of \cref{fig_neuron} physically implemented by the transistor MN\textsubscript{K}.
Only when a sufficient number of input spikes have been injected, so that the critical charge has been reached and $\vM$ has exceeded $\vMth$, an action potential is \emph{fired} by the circuit.
If the input spike train is periodically sustained, one again obtains an action potential train, that is a periodic output spike train whose frequency is directly affected by the magnitude and timing parameters of the input spike train.


%
%
%
%
%

\section{Conclusions and Perspectives}
\label{section:Conclusions and Perspectives}

Spiking neurons are nonlinear circuits that are the fundamental building blocks of spiking neural networks.
Ultra-low-voltage CMOS designs can be found in the literature.
We have focused this work on the study of the excitability property of the neuron, illustrated on an architecture implementing a simplified Morris-Lecar neuronal model.
Relying on conventional SPICE transient simulations allowing for a fully data-driven approach, we have explored and established the spiking criterion.
While the critical charge does depend on the excitation current amplitude and hence cannot be considered as an intrinsic property of the neuron, 
the criterion that the membrane potential must exceed a certain threshold (intrinsic to a given neuron design) was shown to be universal.
The knowledge of the effective nonlinear membrane capacitance allows to relate charge and potential thresholds according to \eqref{eq:DQth = CMtotav.DvMth}.
When a biologically-realistic input synaptic current spike train is supplied, the studied neuron exhibits a typical leaky integrate-and-fire behaviour whose output is an action potential spike train.
Drawing on our experience in SRAM dynamics modelling, the study of the interplay between analog neuron excitability and noise will be part of future work, notably assessing the propagation of spike jitter.



\section*{Acknowledgment}

The authors would like to thank Prof. Rodolphe Sepulchre; Prof. Michele Bonnin, Prof. Alon Ascoli and Prof. Fernando Corinto; Dr. Haralampos Stratigopoulos; Prof. Pietro Maris Ferreira; Prof. Elisabetta Chicca and Mr. Giuseppe Leo for the valuable 
discussions and relevant references that contributed to this work and open perspectives for future work.



\bibliographystyle{IEEEtran}
\bibliography{IEEEabrv,bib}

\end{document}

%% file: mypackages.tex
\usepackage[utf8]{inputenc}
\usepackage[T1]{fontenc}
\usepackage[english]{babel}

\usepackage{textcomp} 

\usepackage{mathtools}
\usepackage{amssymb} 

\usepackage{bm} 

\usepackage{nccmath}

\usepackage[version=3]{mhchem} 
\usepackage{esint} 
\usepackage{dsfont} 
\usepackage[makeroom]{cancel} 
\usepackage{empheq}
\usepackage{physics}
\usepackage[binary-units=true,per-mode=symbol]{siunitx}

\usepackage{graphicx}
\usepackage{float}

\makeatletter
\let\MYcaption\@makecaption
\makeatother

\usepackage[font=footnotesize]{subcaption}

\makeatletter
\let\@makecaption\MYcaption
\makeatother

\usepackage[export]{adjustbox} 

\usepackage{booktabs}
\usepackage{makecell}
\usepackage{threeparttable}
\usepackage{multirow}
 
\usepackage{enumitem}

\usepackage[dvipsnames]{xcolor}

\usepackage[nameinlink,capitalise,noabbrev]{cleveref}
\crefformat{equation}{(#2#1#3)}
\Crefformat{equation}{Equation (#2#1#3)}
\crefrangeformat{equation}{(#3#1#4) to~(#5#2#6)}
\crefmultiformat{equation}{(#2#1#3)}%
{ and~(#2#1#3)}{, (#2#1#3)}{ and~(#2#1#3)}
\Crefmultiformat{equation}{Equations (#2#1#3)}%
{ and~(#2#1#3)}{, (#2#1#3)}{ and~(#2#1#3)}

\usepackage[noadjust]{cite} 
\newcommand*{\citet}[1]{\cite{#1}}

\usepackage{pgf,tikz,pgfplots}
\pgfplotsset{compat=1.10}
\usetikzlibrary{shapes,arrows,calc,arrows.meta,
patterns,backgrounds,positioning,fit,chains}
\tikzset{every picture/.style=black}
\usepackage{circuitikz}

\tikzset{three sided/.style={
        draw=none,
        append after command={
            [shorten <= -0.5\pgflinewidth]
            ([shift={(-1.5\pgflinewidth,-0.5\pgflinewidth)}]				  \tikzlastnode.north east)
        edge([shift={( 0.5\pgflinewidth,-0.5\pgflinewidth)}]				  \tikzlastnode.north west) 
            ([shift={( 0.5\pgflinewidth,-0.5\pgflinewidth)}]				  \tikzlastnode.north west)
        edge([shift={( 0.5\pgflinewidth,+0.5\pgflinewidth)}]				  \tikzlastnode.south west)            
            ([shift={( 0.5\pgflinewidth,+0.5\pgflinewidth)}]				  \tikzlastnode.south west)
        edge([shift={(-1.0\pgflinewidth,+0.5\pgflinewidth)}]				  \tikzlastnode.south east)
        }
    }
}

\usepackage{psfrag}
\usepackage{pstool} 


%% file: mycommands.tex
\makeatletter
\newcommand{\getfontsize}{\f@size pt}
\makeatother

\newcommand*\diff{\mathop{}\!\mathrm{d}}

\input{mycommands_semiconductor.tex}


\input{mycommands_circuits.tex}

%% file: mycommands_circuits.tex
\newcommand*\VDD{V_{\mathrm{DD}}}

\newcommand*{\DTp}{\Delta T_{\textnormal{p}}}

\newcommand*\VM{V_{\mathrm{M}}}

\newcommand*\iex{i_{\mathrm{ex}}}
\newcommand*\Iex{I_{\mathrm{ex}}}
\newcommand*\vM{v_{\mathrm{m}}}
\newcommand*\vGNA{v_{\mathrm{GNA}}}
\newcommand*\vGK{v_{\mathrm{GK}}}
\renewcommand*\VM{V_{\mathrm{m}}}
\newcommand*\VGNA{V_{\mathrm{GNA}}}
\newcommand*\VGK{V_{\mathrm{GK}}}

\newcommand*\CM{C_{\mathrm{m}}}
\newcommand*\CK{C_{\mathrm{K}}}

\renewcommand*{\DTp}{\Delta T_{\mathrm{p}}}
\newcommand*\pw{\DTp}
\newcommand*{\DTq}{\Delta T_{\mathrm{q}}}

\newcommand*\vMres{V_{\mathrm{m,res}}}
\newcommand*\vMth{V_{\mathrm{m,th}}}
\newcommand*\DvMth{\Delta V_{\mathrm{m,th}}}

\newcommand*\ffi{f_{1}}
\newcommand*\fNA{f_{\mathrm{NA}}}

\newcommand*\CMtot{C_{\mathrm{m,eff}}}

\newcommand*\pwcrit{\Delta T_{\mathrm{p,th}}}

\newcommand*\CMtotav{\overline{C}_{\mathrm{m,eff}}}

\newcommand*\DQth{\Delta Q_{\mathrm{th}}}

%% file: mycolors.tex
\definecolor{k}{rgb}{0 0 0}
\definecolor{r}{rgb}{1 0 0}
\definecolor{g}{rgb}{0 1 0}
\definecolor{b}{rgb}{0 0 1}
\definecolor{orange}{rgb}{1,0.7,0}
\definecolor{c}{rgb}{0 1 1}
\definecolor{cc}{RGB}{64 224 208}
\definecolor{m}{rgb}{1 0 1}
\definecolor{khaki}{RGB}{128 128 0}
\definecolor{deepskyblue}{RGB}{0 191 255}
\definecolor{darkMagenta}{rgb}{0.5 0 0.5}
\definecolor{chocolateBrown}{RGB}{98 52 18}
\definecolor{lightBrown}{RGB}{189 154 122}
\definecolor{mybrown}{RGB}{127 37 0}
\definecolor{bordeaux}{RGB}{131 41 85}
\definecolor{violet}{RGB}{127 0 255}
\definecolor{myGreen}{RGB}{134,180,44}
\definecolor{gray_gate}{RGB}{211,208,205}
\definecolor{yellow_oxide}{RGB}{244,231,164}
\definecolor{color_mix}{rgb}{0.7510 0.2510 0.2510}

\definecolor{h}{rgb}{0 0 0}

\definecolor{l}{rgb}{0 0 1}